\documentclass[showpacs,twocolumn,floats,superscriptaddress]{revtex4}
\usepackage{graphicx}
\usepackage{amsmath}

\newcommand{\bq}{\begin{equation}}
\newcommand{\ee}{\end{equation}} \newcommand{\fr}[2]{\frac{#1}{#2}}

\begin{document}
\title{
Polarized Electric Current in Semiclassical Transport  with
Spin-Orbit Interaction}
\author{P. G. Silvestrov}
\affiliation{Instituut-Lorentz, Universiteit Leiden, P.O. Box
9506, 2300 RA Leiden, The Netherlands}
 \affiliation{Theoretische
Physik III, Ruhr-Universit{\"a}t Bochum, 44780 Bochum, Germany}
\author{E. G. Mishchenko}
\affiliation{Department of Physics, University of Utah, Salt Lake
City, UT 84112}

\begin{abstract}

Semiclassical solutions of two-dimensional Schr\"{o}dinger
equation with spin-orbit interaction and smooth potential are
considered. In the leading order, spin polarization is in-plane
and follows the evolution of the electron momentum for a given
subband. Out-of-plane spin polarization appears as a quantum
correction, for which an explicit expression is obtained. We
demonstrate how spin-polarized currents can be achieved with the
help of a barrier or quantum point contact open for transmission
only in the lower subband.

\end{abstract}
\pacs{72.25.-b, 73.23.-b, 72.25.Hg, 03.65.Sq
}
 \maketitle

\section{ Introduction}

Achieving spin manipulation in nanodevices by means of electric
fields (without using less selective magnetic fields) represents
the ultimate goal of spintronics. Spin-orbit interaction, which
couples electron momentum to its spin, is one of the most
promising tools for realizing spin-polarized
transport~\cite{Review1,Review2}. Several schemes leading either
to spin accumulation or to polarization of the transmitted current
induced by the spin-orbit interaction have been put forward.
Predictions of electric field induced spin-accumulation at the
boundaries of a sample, which originates from asymmetric
scattering from impurities \cite{DP,Hirsch99} (extrinsic spin-Hall
effect) or from spin-orbit split band structure
\cite{MNZ,Sinova04} (intrinsic effect), has recently reached a
stage of experimental realization \cite{Fse}. In-plane bulk spin
polarization appears in the electric field in two-dimensional
systems with broken inversion symmetry \cite{E}. Spin polarization
in quantum wires with low carrier density has been shown to occur
due to the interfaces of spin-degenerate and spin-split regions
\cite{Governale}. Interfaces between two-dimensional regions with
different spin-orbit splitting have also been used for that
purpose, in the case of a sharp~\cite{Ramaglia,Ionicioiu} or an
arbitrary~\cite{Finkelstein} interface, as was the scattering from
a sample edge \cite{UB,Govorov}. Other proposals include
polarization due to tunneling through a double-barrier structure
\cite{SR,Koga02} and tunneling between two quantum wires
\cite{GBZ}. Reference~\cite{KK} suggested a three-terminal device
with a spin-orbit split central region as a spin filter, which was
numerically tested by Refs.~\cite{Yamamoto} and~\cite{Ohe}.
Reference~\cite{Eto05} pointed to a possibility of generating
spin-polarized currents by utilizing crossings of spin-orbit-split
subbands belonging to different transverse channels. These
proposals are still lacking experimental realization.

In the present paper we suggest a way to polarize electric
currents by passing them through a region where, by increasing the
external electrostatic potential, the upper spin-orbit-split
subband is locally positioned {\it above} the Fermi level. {The
proposed method utilizes electric gating whose effect is two fold:
(i) it completely suppresses transmission via the upper
spin-orbit-split subband, and (ii) it allows transmission only in
a narrow interval of incident angles in the lower subband.} In
contrast to the proposals which advocate strong variations of the
spin-orbit coupling and, thus, rely on strong gate voltages, our
method requires only weak potentials of the order of a few
millivolts (which is a typical scale of the Fermi energy). In
addition, we predict a specific pinch-off behavior of the
conductance, which would allow to detect polarized currents
without actual measurement of spin.

We consider ballistic electron transport in gated two-dimensional
electron gas with the Hamiltonian
 \bq\label{Ham}
H=\fr{p^2}{2m} +\lambda
(p_y\sigma_x-p_x\sigma_y)+\fr{m\lambda^2}{2}+V(x,y).
 \ee
For the sake of simplicity we concentrate on the case of the
"Rashba" spin-orbit interaction (the same method, however, can be
used for more complicated interactions). Construction of
semiclassical solutions of the Schr\"odinger equation with the
Hamiltonian~(\ref{Ham}) follows the reasoning of the conventional
WKB approach~\cite{Lj,Bolte98,Plet03,Culcer04}, which is valid for
a smooth potential, $\hbar |\nabla V|\ll \min(p^3/m, p^2\lambda)$.
The advantages of semiclassics are twofold. First, it allows us to
obtain approximate analytical solutions for otherwise complicated
problems. Second, as we will see, it turns out to be especially
simple to achieve strong polarization of electron transmission in
the semiclassical regime.

The Mexican hat shape of the effective kinetic energy in the case
of spin-orbit interaction leads to a variety of unusual classical
trajectories (see Fig.~2 below), which have never been
investigated before. Our approach employs strong spin-orbit
interaction (or smooth external potential) sufficient to affect
individual electron trajectories, in contrast to previous
semiclassical treatments \cite{chang,zaitsev} which consider
spin-orbit interaction as a perturbation. Still we do not require
the spin-orbit interaction to be comparable with the bulk value of
the Fermi energy. To produce spin-polarized current, it will be
sufficient to make spin-orbit interaction comparable with the
kinetic energy at some particular area of the system, for example,
near the pinch-off of a quantum point contact.

\section{ Semiclassical wave function}

Without the external potential $V$, the electron spectrum consists
of the two subbands, $E_\pm(p_x,p_y)=(p\pm m\lambda)^2/2m$. The
subbands meet at only one point, $p=0$, and the spin in each
subband is always aligned with one of the in-plane directions
perpendicular to the momentum $\vec{p}$. The semiclassical
electron dynamics~\cite{Lj} naturally captures the essential
features of this translationally invariant limit. The classical
motion in each subband is determined by the equations of motion
which follow from the effective Hamiltonian:
 \bq\label{Hameff}
H_{\rm eff}=\fr{(p\pm m\lambda)^2}{2m} +V(x,y).
 \ee
Despite the fact that spin does not appear in this equation, one
can easily construct semiclassical wave functions, which have spin
pointed within the $xy$~plane perpendicular to the momentum:
 \bq\label{wf}
 \psi_{0}=u e^{iS/\hbar} , \ \
 u=\sqrt{\fr{\rho}{ {2p}}}
 \left(\begin{array}{cc} \sqrt{p_y+ip_x}\\
 \pm\sqrt{p_y-ip_x} \end{array}\right).
 \ee
Here the action $S$ is related to the momentum by $\vec{p}=\nabla
S$, and $\rho=u^\dag u$ is the classical density for a family of
classical trajectories corresponding to a given energy $E$. The
action $S$ obeys the classical Hamilton-Jacobi equation, $(|\nabla
S|\pm m\lambda)^2/2m+V=E$. Application of the Hamiltonian
(\ref{Ham}) to the approximate wave function $\psi_{0}$ gives, after
some algebra,
 \begin{eqnarray}\label{proof}
H\psi_{0}=E\psi_{0} -\fr{i\hbar}{2\rho} \left( \nabla\cdot \rho
\vec{v} \right)\psi_{0} +\hbar\lambda F \sigma_z\psi_{0} ,
 \end{eqnarray}
 where (summation over repeating indices is assumed)
 \bq\label{F}
 F= \fr{
p \mp m\lambda}{2m\lambda p^3}(p_y p_i\partial_i p_x-p_x
p_i\partial_i p_y)
\pm\fr{p_y\partial_x\rho-p_x\partial_y\rho}{2p\rho}.
 \ee
The second term in the rhs of Eq.~(\ref{proof}) vanishes due to
the continuity equation
 \bq\label{continuity}
\nabla \cdot \rho\vec{v}=0 , \ \
\vec{v}={\vec{p}}/{m}\pm\lambda{\vec{p}}/{p}.
 \ee
The last ($\sim\sigma_z$) term in (\ref{proof}) indicates that the
spin of an accelerated electron cannot exactly stay in the plane
of propagation and acquires a small $\sim\hbar\nabla V$ projection
onto the $z$~axis. To take into account this out-of-plain spin
precession one has to go beyond the approximation of
Eq.~(\ref{wf}), which is done by
 \bq\label{wfsz}
\psi= (1+\hbar f\sigma_z)\psi_{0}.
 \ee
Since
$(H-E)f\sigma_z\psi_{0} = \mp 2\lambda p f\sigma_z\psi_{0}$,
to the lowest order in $\hbar$,
one can relate the functions $F$ 
and $f$
 \bq
f=\pm {F}/{2 p},
 \ee
and  find the out-of-plane spin density [$F$ is found from
Eq.~(\ref{F})]
 \bq\label{sigmaz}
\psi^\dagger\sigma_z\psi =\pm\fr{\hbar\rho}{2p}F.
 \ee

Note that Eq.~(\ref{sigmaz}) does not describe the nonadiabatic
transitions between subbands. After the electron leaves the region
with nonzero potential gradient, $\nabla V\neq 0$, the in-plain
spin orientation is restored.

The out-of-plane polarization of the electron flow in the external
potential is a subject of the rapidly developing field of the
spin-Hall effect~\cite{DP,Hirsch99,MNZ,Sinova04,Fse}. Our result,
Eqs.~(\ref{F}) and~(\ref{sigmaz}), incorporates previous
calculations of Ref.~\cite{Sinova04} which were restricted to the
one-dimensional form of the potential, $V(x)$, with $p_y$ being
the integral of motion. The validity of Eq.~(\ref{sigmaz}),
however, is not restricted to a simple one-dimensional case and
describes the out-of-plain polarization for {\it any} smooth
two-dimensional potential (including confining potentials which
create quantum wires, quantum dots, etc.). In particular,
Eq.~(\ref{sigmaz}) may serve as a good starting point for an
analytical calculation of the edge spin accumulation in ballistic
quantum wires~\cite{Brogumilovich,Balsejro}. We leave further
investigation of these interesting effects for subsequent
research.

Solutions of the form, Eq.~(\ref{wf}), have clear and important
consequences. During its motion, an electron changes the momentum
$p$ but always remains in the same spin-subband. To change the
subband the electron trajectory should pass through the degeneracy
point where both components of momentum vanish simultaneously,
$\vec{p}=0$, which is generically impossible. Moreover, with the
proper use of potential barriers, one may realize a situation where
electrons {\it of only one subband} are transmitted and the others
are totally reflected. This leads to strong polarization of the
transmitted electron flow.

\begin{figure}
\includegraphics[width=7.3cm]{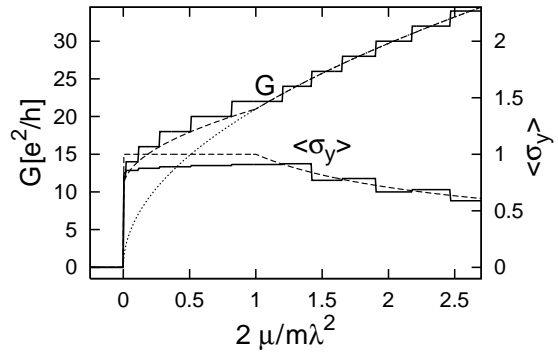}
\caption{ Conductance (in units of $e^2/h$), and spin polarization
of the current vs gate voltage (in units of $m\lambda^2/2$).
Dashed lines show the smoothed curves (\ref{Landauer},\ref{Spin}),
solid lines show the quantized values for $m\lambda
L/\hbar=10.5\pi$. Dotted line shows the conductance without
spin-orbit interaction. }
\end{figure}

\section{ Sharvin conductance}

To give an example of such a spin-polarized current let us
consider transmission through a barrier, $V(x)$, varying along the
direction of a current propagation. We assume periodic boundary
conditions in the perpendicular direction ($y+L\equiv y$). As such
a condition makes $p_y$ the integral of motion, mixing of orbital
channels, which is strongly suppressed for generic smooth
potential~(\ref{Hameff}), is now absent exactly. For a smooth
potential $V(x)$ the conduction channels may either be perfectly
transmitting or completely closed. The conserved transverse
momentum takes the quantized values, $p^{n}_{y}=2\pi\hbar n/L$.
Consider the functions
 \bq
E^{n}_\pm(p_x)=\fr{(p^n\pm m\lambda)^2}{2m}  , \ \
p^n=\sqrt{p_x^2+{p^{n}_{y}}^2}
 \ee
For $n\neq 0$ the function $E^{n}_\pm(p_x)$
splits into two distinct branches. At any point
$x$ the equation
 \bq\label{channel}
E^{n}_\pm(p_x) = E_F -V(x)
 \ee
yields solutions $p_x^L$ and $p_x^R$, corresponding to left- and
right-moving electrons. Application of a small bias implies, e.g.,
the excess of right movers over left movers far to the left from
the barrier. Particles are transmitted freely above the barrier if
Eq.~(\ref{channel}) has a solution, $p_x^R$, for any $x$. Let
$\mu=E_F-V_{\max}$ be the difference between the Fermi energy and
the maximum of the potential. The $n$th channel in the upper
branch opens when
 \bq\label{muplus}
\mu=\left({2\pi\hbar |n|}+{m\lambda L}
\right)^2/2mL^2.
 \ee
For the lower branch $E_-^{n}(p_x)$ Eq.~(\ref{channel}) has four
solutions (two for right and two for left movers) for $|n|<
m\lambda L/2\pi\hbar$ and $x$ close to the top of the barrier.
However, far from the barrier (where the excess of right-movers is
created) there are still only two crossings described by
Eq.~(\ref{channel}), one for right and one for left movers. As a
result, all the extra electrons injected at $x=-\infty$ follow the
evolution of a solution of Eq.~(\ref{channel}) with the largest
positive $p_x$. For all $|n|< m\lambda L/2\pi\hbar$ such a
solution does exist for any positive $\mu$. Thus, at $\mu=0$ as
many as $n_0=m\lambda L/\pi\hbar$ channels {\it open up
simultaneously}. The channels with higher values $|n|>m\lambda
L/2\pi\hbar$ in the lower subband $E_-^{n}$ open when
 \bq\label{muminus}
\mu=\left({2\pi\hbar |n|}-{m\lambda L}
\right)^2/2mL^2.
 \ee
According to the Landauer formula, ballistic conductance is given
by the total number of open channels multiplied by the conductance
quantum $G_0=e^2/h$
 \bq\label{Landauer}
G=G_0\fr{ L}{\pi\hbar} \left\{\begin{array}{cl}
 \sqrt{{2\mu m}}+m\lambda,~~
& 0<\mu <m\lambda^2/2 \\
 2 \sqrt{{2m\mu}}, &
\mu >m\lambda^2/2. \end{array}\right.
 \ee
This dependence $G(\mu)$ is shown in Fig.~1. The striking evidence
of the presence of spin-orbit interaction is the huge jump of the
conductance at the pinch-off point, as opposed to the conventional
square-root increase in the absence of spin-orbit coupling. This
jump is a consequence of the ``Mexican-hat'' shape of the spectrum
$E_-(p_x,p_y)$. Accuracy of Eqs.~(\ref{muplus})
and~(\ref{muminus}) is sufficient to resolve the steps in the
conductance due to the discrete values of $|n|=0,1,2,...,$
(conductance quantization), as shown in Fig.~1. The steps in
$G(\mu)$ are abrupt in the limit $dV/dx\rightarrow 0$.

Close to the pinch-off, at $\mu\lesssim m\lambda^2$, the conserved
$p_y$ component of the electronic momentum varies for different
transmitted channels within the range $|p_y|\lesssim m\lambda$.
Therefore, far from the barrier, where the Fermi momentum is large
$p_F \gg m\lambda$,  we have $p_x\gg p_y$ and transmitted
electrons propagate in a very narrow angle interval
$|\theta|<\sqrt{m\lambda^2/2E_F}\ll 1$. Since the electron spin is
perpendicular to its momentum, we conclude that the current due to
electrons from each of the subbands is almost fully polarized. The
total polarization of the transmitted current is given by the
difference of two currents
 \bq\label{Spin}
\langle{\sigma_y}\rangle= {\langle\psi^\dagger{\sigma_y}v_x\psi
\rangle}/{\langle\psi^\dagger v_x\psi\rangle }= \min
(1,\sqrt{{m\lambda^2}/{2\mu}}),
 \ee
which is also depicted in Fig.~1. This current polarization may
also be viewed as a creation of in-plain nonequilibrium spin
density, maximal on the barrier.

Vanishing transmission for electrons from the upper band for $0<\mu
< m\lambda^2/2$~(\ref{Landauer}) resembles the total internal
reflection suggested for creation of polarized electron beams in
Ref.~\cite{Finkelstein}. Unlike the latter case, in our proposal
there is no need to collimate incident electron flow, since the
upper band electrons are reflected at any angle.

Semiclassical formulas (\ref{Landauer}) and~(\ref{Spin}) are valid
provided that there are many open transmission channels, and
account correctly for the electrons with $p^{n}_y\neq 0$. The case
$n=0$, however, requires special attention. The curve
$E^{0}_\pm(p_x)$ does not split into the lower and upper branches,
but instead consists of two crossing parabolas shifted
horizontally. Right movers from both parabolas are transmitted or
reflected simultaneously. The electron flow due to the channels
with $n=0$ is, therefore, unpolarized.  For small $n\neq 0$ the
crossing of two parabolas is avoided. However, the electrons from
the upper subband $E_{+}^{n}$ may tunnel into the lower branch
$E_{-}^{n}$ in the vicinity of the point $p_x=0$, which results in
the decrease of spin-polarization of the current. Let the barrier
near the top has a form $V(x)=-m\Omega^2 x^2/2$. Simple estimation
shows that classically forbidden transition between the subbands
do not change the net polarization of the current as long as
$\hbar\Omega \ll m\lambda^2$.

Our results Eqs.~(\ref{Landauer}) and~(\ref{Spin}) were obtained
for the periodic boundary conditions. However, the boundary
conditions do not play important role for the conductance
($G\propto L$) if the width of the "wire" is large compared with
the width of the barrier, i.e., if
$L\gg\sqrt{\hbar/m\Omega}\gg\hbar/m\lambda$. If the transverse
confinement in the wide wire is ensured by the smooth
potential~\cite{endnonote} the semiclassical transmitted
scattering states may be constructed explicitly using the method
of Ref.~\cite{Sil03}. However, since the spin-orbit interaction in
our approach appears already in the classical
Hamiltonian~(\ref{Hameff}), calculation of smoothed
conductance~(\ref{Landauer}) requires only a simple counting of
classical trajectories~\cite{endnote}. Our next example below
demonstrates such semiclassical treatment of realistic boundary
conditions.

\section{ Quantum Point Contact}

Let us consider probably the most experimentally relevant example
of a quantum point contact, described by the potential
 \bq\label{VQPC}
V(x,y)= -\fr{m\Omega^2x^2}{2} +\fr{m\omega^2 y^2}{2}.
 \ee
We will see that even in this simple model the electron flow in
the presence of spin-orbit interaction acquires a number of
interesting and peculiar features. Classical equations of motion
follow in the usual manner from the effective Hamiltonian
(\ref{Hameff}): $\dot{\vec{r}}=\partial H_{\rm eff}/\partial
\vec{p}$, $\dot{\vec{p}}=-\partial H_{\rm eff}/\partial \vec{r}$.
We consider quantum point contact~(QPC) close to the  opening with
only the lower $E_-$ subband contributing to the conductance. A
crucial property of the Hamiltonian $H_{\rm eff}$,
Eq.~(\ref{Hameff}), is the existence of a circle of minima of the
kinetic energy at $|p|=m\lambda$. Expanding around a point on this
circle, $p_{x_0}=m\lambda\cos\alpha$,
$p_{y_0}=m\lambda\sin\alpha$, one readily finds the equations of
motion for ${\cal P} =p_x\cos\alpha+p_y\sin\alpha- m\lambda\ll
m\lambda$,
 \bq\label{motimotion}
 \ddot{\cal P}+(-\Omega^2\cos\alpha^2+\omega^2\sin\alpha^2){\cal P}=0
 \ , \
 \dot{\alpha}=0.
 \ee
The trajectory is found from the relations, $\dot{x}={\cal
P}\cos\alpha/m \ , \ \dot{y}={\cal P}\sin\alpha/m$. We observe
from Eq.~(\ref{motimotion}) that only the trajectories within the
angle
 \bq\label{tanalpha}
\tan|\alpha|< \tan\alpha_0={\Omega}/{\omega}
 \ee
are transmitted through QPC. Trajectories with larger angles are
trapped (oscillate) within the point contact. Examples of both
types of trajectories are presented in Fig.~2. Quantization of
trapped trajectories would give rise to a set of (extremely)
narrow resonances in the conductance, specific for spin-orbit
interaction. {We leave the detailed investigation of these narrow
features for future research. Below we consider only the smoothed
conductance.}

To calculate the current $J$ through QPC one has to integrate over
the phase space of the states which are transmitted from left to
right,
 \bq
J=\int\limits dy\int\limits ev_x\fr{d^2p}{(2\pi\hbar)^2}=G{\cal
V},
 \ee
and have the energy within the interval $\mu-e{\cal V}/2< E_{-}<
\mu+e{\cal V}$/2, with ${\cal V}$ standing for the applied
voltage. In this section we define $\mu$ as the difference between
the Fermi energy and the value of the potential at the saddle
point $\mu=E_F-V(0,0)$. The integral is most simply evaluated at
$x=0$ (with the velocity given by $v_x={\cal P}\cos\alpha /m$).
The allowed absolute values of the momentum are
 \bq
2\mu-{e{\cal V}}-{m\omega^2y^2} < {{\cal P}^2}/{m}< 2\mu+{e{\cal
V}}-{m\omega^2 y^2}.
 \ee
The angle interval of transmitting trajectories consists of two
domains: $|\alpha|<\alpha_0,~{\cal P}>0$, and
$|\alpha-\pi|<\alpha_0,~ {\cal P}<0$. The appearance of the latter
range of integration is highly non-trivial. A simple reasoning
shows that the particles with the velocity antiparallel to the
momentum ($v_x>0$, $p_x<0$) should not contribute to the
conduction in the case of a transition through a one-dimensional
barrier $V=V(x)$, see Eq.~(\ref{Landauer}). Despite corresponding
to the right-moving electrons, these states {\it do not originate}
in the left lead. Indeed, they exist only in the vicinity of
$x=0$, but disappear as $x\rightarrow -\infty$ and, thus, cannot
be populated by the excess electrons (except due to the tunneling
transitions which are irrelevant in the semiclassical regime).
Such trajectories, however, {\it do exist} in QPC,
Eq.~(\ref{VQPC}), as demonstrated in Fig.~2. After passing through
QPC the trajectory bounces at the wall reversing its velocity.
This kind of classical turning points, where both components of
the velocity vanish simultaneously, are specific for the effective
Hamiltonian~(\ref{Hameff}). The existence of transmitting
trajectories with $|\alpha-\pi|<\alpha_0,~ \varrho<0$ results in
the doubling of the conductance. Simple calculation yields
 \bq\label{conduc}
G=G_0\fr{4m\lambda\sin\alpha_0} {\pi\hbar\omega}
\sqrt{\fr{2\mu}{m}}.
 \ee
The presence of a threshold angle $\alpha_0$, as well as the
square-root dependence of $G(\mu)$, are in a sharp contrast to the
well-known result $G=G_0\mu/\pi\hbar\omega$, in the absence of
spin-orbit interaction.

Equation (\ref{conduc}) is valid in the case of many open
channels. Since Eq.~(\ref{motimotion}) describes only the
linearized electron dynamics, Eq.~(\ref{conduc}) is formally valid
if $\mu\ll m\lambda^2$. Nevertheless, the current remains totally
polarized for $0<\mu < m\lambda^2/2$ [similar to Eq.~(\ref{Spin})]
 \bq\label{SpinQPC}
\langle{\sigma_y}\rangle= {\langle\psi^\dagger{\sigma_y}v_x\psi
\rangle}/{\langle\psi^\dagger v_x\psi\rangle }= 1.
 \ee
With increasing the chemical potential, $\mu > m\lambda^2/2$,
transmission via the upper subband  $E_{+}$ kicks in and the
degree of polarization gradually decreases, {similarly to
Eq.~(\ref{Spin}), though with different, more complicated,
dependence of spin-polarization on $\mu$}. Note that transmission
of different orbital channels through QPC is independent as long
as the confining potential (\ref{VQPC}) is smooth over a distance
of the characteristic spin-orbit length $\hbar/ m\lambda$. It is
easy to see that this requirement is equivalent to the condition
that $(\omega, \Omega) \ll m\lambda^2/\hbar$. This is also a
condition of large conductance $G\gg G_0$.

\begin{figure}
\includegraphics[width=7.cm]{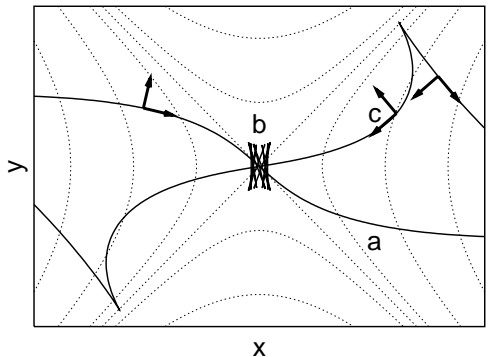}
\caption{ Three kinds of trajectories in the point contact. $a$,
transmitted trajectory whose momentum is always collinear with the
velocity. $b$, trajectory bouncing inside the QPC. This trajectory
is periodic in the linearized approximation described in the text,
while the exact calculation for finite amplitude shows its slow
drift. $c$, transmitted trajectory whose momentum inside the
contact is opposite to the velocity. Electrons flow from left to
right. Arrows show momentum and spin orientations. Few
equipotential lines are also shown. }
\end{figure}

\section{ Discussion}

 In both analyzed systems (of ballistic Sharvin conductance and of
QPC) polarization of current is achieved when many channels are
transmitting. As a consequence of the Kramers degeneracy,
transmission eigenvalues always appear in pairs in the presence of
time-reversal symmetry, leading to the prohibition of the
spin-current in the lowest ($n=0$) conducting channels
(cf.~Ref.~\cite{KK}). In the case of higher channels, however, the
degenerate transmission eigenvalues belong to the same spin-orbit
subband and carry, respectively, the same spin polarization. For
example, in the case of the QPC any transmitted trajectory
$x(t),y(t)$ (e.g., one of the two shown in Fig.~2) is accompanied
by its mirror reflection $x(t),-y(t)$ with identical transmission.

In InAs-based heterostructures, typical value of spin-orbit coupling
\cite{Grun} is $\lambda\hbar =2\times 10^{-11}eVm$. Characteristic
spin-orbit length $l_R=\hbar/m^*\lambda = 100 $ nm and energy
$m^*\lambda^2/2= 0.1$ meV. In order to have strongly spin-polarizing
QPC, the latter should support many transmitting channels at
chemical potential $\mu\sim m^*\lambda^2/2 \gg \hbar\omega$. This
condition can, equivalently, be written in terms of the width of the
point contact $\Delta y$, see Eq.~(\ref{VQPC}), as $\Delta y \gg
l_R$. This is a realistic condition for typical ballistic
constrictions.

To conclude, we have proposed a way to polarize currents in the
ballistic regime by means of using electric gates to suppress
transmission in the upper spin-orbit-split subband. The
polarization is stronger when there are many transmitting channels
in the lower subband. This is exactly the condition when the
semiclassical expansion in powers of $\hbar$ is applicable. An
obvious advantage of our scheme is that we do not require the
spatial modulation of the strength of spin-orbit interaction.
Neither do we need a restricted angle of incident electrons in
order to have a polarized current.

\begin{acknowledgments}

We have benefited from discussions with G.E.W.~Bauer,
C.W.J.~Beenakker, and B.I.~Halperin. This work was supported by
the Dutch Science Foundation NWO/FOM, by the SFB TR 12, and by
the DOE, Office of Basic Energy Sciences, Award
No.~DEFG02-06ER46313.
\end{acknowledgments}


\begin{thebibliography}{99}

\bibitem{Review1} {\it Semiconductor Spintronics and Quantum Computation},
edited by D.D.~Awschalom, D.~Loss, and N.~Samaranth (Springer,
Berlin, 2002).

\bibitem{Review2} I.~Zutic, J.~Fabian, and S.~Das~Sarma, Rev. Mod. Phys. {\bf
76}, 323 (2004).

\bibitem{DP} M.I.~D'yakonov, V.I.~Perel, Phys.\ Lett.\ A {\bf 35},
459 (1971).

\bibitem{Hirsch99} J.E.~Hirsch, Phys. Rev. Lett.\ {\bf 83}, 1834 (1999).

\bibitem{MNZ} S.~Murakami, N.~Nagaosa, and S.-C.~Zhang, Science {\bf
301}, 1348 (2003); Phys.\ Rev.\ B {\bf 69}, 235206 (2004).

\bibitem{Sinova04} J.~Sinova, 
D.~Culcer, Q.~Niu, N.A.~Sinitsyn, T.~Jungwirth, and
A.H.~MacDonald, Phys.\ Rev.\ Lett.\ {\bf 92}, 126603 (2004).

\bibitem{Fse}
Observation has been reported by Y.K.~Kato, 
R.C.~Myer, A.C.~Gossard, and D.D~Awschalom, Science {\bf 306},
1910 (2004); J.~Wunderlich, B.~Kastner, J.~Sinova, and
T.~Jungwirth, Phys.\ Rev.\ Lett.\ {\bf 94}, 047204 (2005);
S.O.~Valenzuela and M.~Tinkham, Nature {\bf 442}, 176 (2006); vast
number of theoretical contributions to spin-Hall effect is far
beyond the scope of our paper.

\bibitem{E} V.M.~Edelstein, Solid\ State\ Commun.\ {\bf 73}, 233 (1990).

\bibitem{Governale} M.~Governale and U.~Z\"{u}licke, Phys. Rev. B
{\bf 66}, 073311 (2002).

\bibitem{Ramaglia} V.~M. Ramaglia, 
D.~Bercioux, V.~Cataudella, G.~De~Fillips, C.A.~Perroni, and
F.~Ventriglia, Eur.~Phys.~J.~B~{\bf 36}, 365 (2003);
V.~M.~Ramaglia, D.~Bercioux, V.~Cataudella, G.~De~Fillips, and
C.A.~Perroni, J.~Phys.~Condens.~Matter~{\bf 16}, 9143 (2004).

\bibitem{Ionicioiu} R.~Ionicioiu and I.~D'Amico, Phys. Rev. B
{\bf 67}, 041307(R) (2003).

\bibitem{Finkelstein} M.~Khodas, A.~Shekhter, and A.M.~Finkel'stein,
Phys. Rev. Lett. {\bf 92}, 086602 (2004).

\bibitem{UB} G.~Usaj and C.A.~Balseiro, Phys. Rev. B {\bf 70},
041301(R) (2004).

\bibitem{Govorov} A.O.~Govorov, A.V.Kalameitsev, and J.P.~Dulka,
Phys. Rev. B {\bf 70}, 245310 (2004).

\bibitem{SR} E.A. de Andrada~e~Silva and G.C.L.~Rocca,
Phys. Rev. B {\bf 59}, R15583
(1999).

\bibitem{Koga02} T.~Koga, J.~Nitta, H.~Takayanagi, and S.~Datta,
Phys. Rev. Lett. {\bf
88}, 126601 (2002).

\bibitem{GBZ} M.~Governale, D.~Boese, U.~Z\"ulicke, and C.~Schroll,
Phys. Rev. B {\bf 65}, 140403(R) (2002).

\bibitem{KK} A.A.~Kiselev and K.W.~Kim, Appl.\ Phys.\ Lett.\ {\bf 78}, 778
(2001).

\bibitem{Yamamoto} M.~Yamamoto, T.~Ohtsuki, and B.~Kramer,
Phys. Rev. B {\bf 72}, 115321 (2005).

\bibitem{Ohe} J.I.~Ohe, M.~Yamamoto, T.~Ohtsuki, and J.~Nitta,
Phys. Rev. B {\bf 72}, 041308(R) (2005).

\bibitem{Eto05}  M.~Eto, T.~Hayashi, and Y.~Kurotani, J. Phys. Soc. Jpn.,
{\bf 74}, 1934 (2005).

\bibitem{Lj} R.G.~Littlejohn and W.G.Flynn, Phys. Rev. A {\bf 44}, 5239
(1991); {\bf 45}, 7697 (1992).

\bibitem{Bolte98} J.~Bolte and
S.~Keppeler, Phys. Rev. Lett. {\bf 81}, 1987 (1998).

\bibitem{Plet03} M.~Pletyukhov and O.~Zaitsev, J. Phys. A: Math. Gen. {\bf 36},
5181 (2003).

\bibitem{Culcer04} D.~Culcer, J.~Sinova, N.~A.~Sinitsyn, T.~Jungwirth,
A.~H.~MacDonald, and Q.~Niu, Phys. Rev. Lett. {\bf 93}, 046602
(2004).

\bibitem{chang} C.-H. Chang, A. G. Mal'shukov, and K. A. Chao, Phys.
Rev. B {\bf 70}, 245309 (2004).

\bibitem{zaitsev} O. Zaitsev, D. Frustaglia, and K. Richter, Phys.
Rev. Lett. {\bf 94}, 026809 (2005); Phys. Rev. {\bf 72}, 155325
(2005).

\bibitem{Brogumilovich} B.~K. Nikolic, S. Souma, L.~P.~Zarbo,
J.~Sinova, Phys. Rev. Lett. {\bf 95}, 046601 (2005).


\bibitem{Balsejro} G.~Usaj and C.A.~Balseiro, Europhys.\ Lett.\ {\bf 72},
631 (2005).

\bibitem{endnonote} Once the semiclassical treatment of the
Hamiltonian (\ref{Ham}) is established, Eqs.~(\ref{Hameff}) and
(\ref{wf}), one can use classical trajectories for the calculation
of conductance \cite{Beenakk91}. Similar calculations would also
give the in-plane spin polarization, since in our case spin is
completely determined by the ``classical'' subband index. Thus,
even for a wide wire, sufficiently smooth boundaries allow us to
avoid a discussion of the chanel mixing, which is crucial for the
perturbative treatment of the spin-orbit interaction in ballistic
wires~\cite{Kirczen01}. Numerically, the nonperturbative chanel
mixing may be investigated by means of the recursive Green's
function method (see e.g. Ref.~\cite{Zhu99}), although this
calculation may not be as simple in our case of multichannel wire
with many channels having exponentially weak transmission.

\bibitem{Beenakk91} C.W.J. Beenakker and H. van Houten,
Phys. Rev. B {\bf 43}, 12066 (1991).

\bibitem{Kirczen01} F. Mireles and G. Kirczenow,
Phys. Rev. B {\bf 64}, 024426 (2001).

\bibitem{Zhu99} S.-L. Zhu, Z.D. Wang, and L. Hu, J. Appl. Phys.
{\bf 91}, 6545 (2002).

\bibitem{Sil03} P.G. Silvestrov, M.C. Goorden, and C.W.J. Beenakker,
Phys. Rev. B {\bf 67}, 241301(R) (2003).

\bibitem{endnote}
In the case of hard wall boundaries scattering at the boundary
will lead to electron transition between spin-orbit-split
subbands. A certain amount of "wrong" spin polarization may in
this case be transferred through the barrier in a form of
evanescent (decaying towards the center of the wire) modes. Still,
if the width of the wire is large compared to the width of the
barrier a fraction of electronic trajectories which hit the
boundary while flying above the barrier is small and
Eqs.~(\ref{Landauer}) and (\ref{Spin}) remain valid. Electrons
transferred through the central region of the barrier will also
eventually hit the walls, leading to the relaxation of
spin-polarization of the current. However, since the barrier
strongly collimates the transmitted electron angles the spin
polarization survives at large distances $\sim
L\sqrt{E_F/m\lambda^2}$ after(before) the barrier.

\bibitem{Grun} D. Grundler, Phys. Rev. Lett. {\bf 84}, 6074 (2000).

\end{thebibliography}
\end{document}